\documentclass[aps, prb,
twocolumn,amsmath,amssymb,showpacs,groupedaddress]{revtex4}
\usepackage{graphicx}
\graphicspath{{../}}
\usepackage{epstopdf}
\begin{document}

\preprint{}

\title{Comment on "Towards a bulk theory of flexoelectricity"}

\author{Alexander K. Tagantsev}
\affiliation{Ceramics Laboratory, Swiss Federal Institute of Technology (EPFL), CH-1015 Lausanne, Switzerland}
\date{\today}

\pacs{77.22.-d, 77.65.-j, 77.90.+k}

\maketitle

The flexoelectric effect consists of linear response of the dielectric polarization to a strain gradient.
It was shown long ago \cite{Tag1985} that, in crystalline solids, this effect is controlled by 4 mechanisms of different  physical nature: (i) static bulk ferroelectricity, (ii) dynamic bulk ferroelectricity, (iii) surface piezoelectricity, and (iv) surface flexoelectricity.
The contributions of these mechanisms can be comparable in size.
There exist situations where this effect manifests itself differently: first, as a polarization wave following an acoustic wave (e.g. acoustic phonon) and, second, when the electromechanical response of the  material is tested by using inhomogeneously deformed parallel-plate short-circuited capacitor (e.g. bent).
It was shown that in the first case mechanisms (i) and (ii) are active whereas, in the second case, mechanisms (i), (iii), and (iv) are.
This is in contrast to the piezoelectric effect which is controlled by a unique mechanism.

In his resent publication, for a specific model, Resta \cite{Resta} performed calculations for a part of the contribution of mechanism (i) (static bulk ferroelectricity).
Based on the results obtained  he came to a conclusion that the flexoelectric effect is controlled by a unique mechanism (similar to the piezoelectric effect).
Below  I will show that this conclusion is wrong and it originates only from the implicit approximations done by this author and specifics of his model.

\emph{Dynamic bulk ferroelectricity (ii).}
Resta has attempted to show that the polarization in the acoustic wave is related to the strain gradient identically to the case of a homogeneous strain gradient.
This was done on the lines of the treatment of the piezoelectric effect by Martin \cite{Martin} who compared the polarization response to static sinusoidal modulation of elastic strain $u =u_0\exp(ikx)$ (where $k$ is the wave vector and $x$ is the distance) with that in the homogeneous situation.
Such method is fully justified for the piezoelectric effect since this effect is a first-order spatial-dispersion effect (i.e. the amplitude of the polarization wave is proportional to $ka$, where $a$ is the lattice constant) whereas the frequency dispersion effects (for the non-dissipative lattice dynamics) reveal themselves only in contributions proportional to the even powers of frequency $\omega$, starting from $\omega^2$.
However, this method does not work for flexoelectricity where the amplitude of the polarization wave is proportional to $(ka)^2$  whereas the frequency dispersion effects cannot be neglected since the $\omega^2-$ and $(ka)^2-$ contributions are of the same order of magnitude \cite{Tag1985}.
An additional shortcoming of Resta's paper is the choice of the model for the discussion of the dynamic effects in flexoelectricity: calculations were performed for monatomic crystal, neglecting the contribution of internal strains, whereas it is known  \cite{Tag1985} that this effect occurs in case where the masses of the atoms in the unit cell are different, being controlled by to the internal strains.

\emph{Surface piezoelectricity (iii).} The contribution of the surface piezoelectricity to the total flexoelectric response in the case of homogeneous strain gradient in a finite sample is related to the distorting (inversion symmetry braking) effect of the surface of the sample on the crystalline lattice.
Due to this effect, thin (possibly, atomically thin) surface-adjacent piezoelectric layers form,
the effective piezoelectric modulus of the layers at the opposite faces of the crystalline plate having opposite signs.
Once inhomogeneously deformed (e.g. bent), the strain difference at the the opposite faces of the plate leads to a difference in the absolute values of the dipole moments generated in the corresponding surface-adjacent layers.
Being of the opposite signs, these dipole moments yield  the total polarization response proportional to the strain gradient applied, which can be of the order of the static bulk flexoelectricity \cite{Tag1985}.
The lattice distortions caused  by  the surface (the driving force of the surface piezoelectricity contribution to the flexoelectric effect) have been neglected in Resta's calculations.

\emph{Surface flexoelectricity (iv).} This is the second surface contribution to the flexoelectric response with a contribution comparable to that of the surface piezoelectricity.
Thus, its presence is not essential for the quantitative picture of the phenomenon questioned by Resta (i.e. a possibility of an essential surface contribution).
Nevertheless, demonstrating its presence or absence is of a scientific interest.
In Ref. \cite{Tag1985}, I have shown that applying an approach,  earlier used for elimination of spurious (of the surface origin) contributions to the piezoelectric response,  one cannot eliminate the surface contribution to the flexoelectric effect.
The surface contribution appeared this way was expressed in terms of the trace of the quadruple moment tensor, $I$, of all charges of the system (including electronic and ionic charges participating in the screening of the macroscopical electric field in the system).
Resta argus that one can show that $I=0$ using results from other papers, though no calculations are presented.
Beyond any doubts, a dedicated microscopic calculation of $I$ for any realistic periodic crystalline structure with compensating charges is welcome in order to know more about the surface flexoelectricity.

Work supported by Swiss National Science Foundation.

\end{document}